\documentclass[10pt]{revtex4}
\usepackage{amssymb,amsmath}
\usepackage{latexsym}
\usepackage{color,paralist}
\usepackage[normalem]{ulem}
\usepackage{suffix}
\usepackage{epsfig}

\DeclareMathOperator{\Tr}{Tr}
\DeclareMathOperator{\tr}{tr}

\newcommand\VTT[1]{\text{{\color[rgb]{1,0,0}~[#1]}}}
\WithSuffix\newcommand\VTT*[1]{{{\color[rgb]{1,0,0}[#1]}}}
\WithSuffix\newcommand\VTT+[1]{{{\color[rgb]{1,0,0}[#1]\\}}}
\WithSuffix\newcommand\VTT@[1]{{{\color[rgb]{1,0,0}[#1]\\\\}}}

\definecolor{mysquiggly}{rgb}{0.9,0,0}

\makeatletter
\def\squiggly{\bgroup \markoverwith{\textcolor{mysquiggly}{\lower3.5\p@\hbox{{\sixly \char58}}}}\ULon}
\makeatother

\allowdisplaybreaks

\begin{document}

\title{ Inflationary universe from  anomaly-free $F(R)$-gravity  }

\author{ Aroonkumar Beesham$^{1,2}$\footnote{abeesham@yahoo.com} and    Kazuharu Bamba$^{3}$\footnote{bamba@sss.fukushima-u.ac.jp}   }
\address{    $^1$Department of Mathematical Sciences, University of Zululand, P Bag X1001, Kwa-Dlangezwa 3886, South Africa \\ $^2$Faculty of Natural Sciences, Mangosuthu University of Technology, Jacobs 4026, South Africa \\ $^3$Division of Human Support System, Faculty of Symbiotic Systems Science, Fukushima University, Fukushima 960-1296, Japan 
}

\begin{abstract}
By adding a three dimensional manifold to an eleven dimensional manifold in supergravity, we obtain the  action of $F(R)$-gravity  and find that it is anomaly free. We calculate the scale factor of the inflationaty universe in this model, and observe that it is related to  the slow-roll parameters. The scalar-tensor ratio  R\_(scalar-tensor) is in good agreement with experimental data.


 \end{abstract}
 \date{\today}

\maketitle
\section{Introduction}

One of the best models for
considering an inflationary universe is suggested by modified theories of gravity like $f(R)$
gravity, whose Lagrangian is a function $f(R)$ of the scalar curvature $R$ in a 4-dimensional universe \cite{p1,p2,p3,p4,p5}. The special property of these theories is the existence of the effective cosmological constant  in such a way that early-time inflation and late-time cosmic acceleration are naturally unified within a single theory \cite{p1}.  Some of these  $f(R)$  gravity theories, especially a power-law mechanism and Yang-Mills $f(R)$  gravity,   give the best fit values compatible with cosmological parameters like the  tensor-to-scalar ratio within the permitted ranges derived by the Planck and BICEP2 observations \cite{p2}. In some other versions of these theories, a dynamical scalar field ($\varphi$) is coupled to gravity ($f(R,\varphi)$) whose dynamics allow for exit from inflation and which gives rise to the correct amount of inflation in agreement with observational data \cite{p3}. Also, in some modified gravity theories, $F$   is a generic function of the curvature scalar  and the Gauss-Bonnet topological parameter. Cosmological dynamical results which are obtained by two effective masses (lengths) correspond to both of these parameters and work, respectively, at early and very early epochs of the evolution of the universe  \cite{p4}. Furthermore, modified teleparallel gravity leads to second order equations and solves the particle horizon problem in a spatially flat cosmic space by providing an initial exponential expansion without resorting to an inflaton particle  \cite{p5}. For more reviews on the connections between inflation, the dark energy problem and 
modified gravity theories see, for instance,~\cite{p6}.

Now, the question that arises is whether $F(R)$ gravity could be anomaly free or not? Also, it is required to consider the role of anomaly cancellation in the evolution of the universe during the inflationary epoch.  Recently, the exact forms of two-dimensional and four-dimensional anomalies from higher derivative gravity and $F(R)$ gravity have been calculated \cite{p7}. On the other hand, the conformal-anomaly driven inflation in $F(R)$  gravity without using the scalar-tensor representation has been studied. It has been shown that in $F(R)$  gravity, the curvature perturbations with their high amplitude  which are consistent with observations, are created during inflation \cite{p8}. Motiviated by these works, we propose a theory which lives on an (11+3)-dimensional manifold with two 11-dimensional manifolds and one 3-dimensional manifold. We will show that our universe is a part of an 11-dimensional manifold which is connected with the other 11-dimensional manifold by an extra 3-dimensional  manifold. Two 11-dimensional manifolds interact with each other via exchanging fields which move along the 3-dimensional manifold. These fields are the main cause for the appearance of $F(R)$-gravity in the four dimensional universe.

This paper consists of two main parts. In section \ref{o1}, we show that by adding 3-dimensional manifold to an 11-dimensional one, $F(R)$-gravity emerges.  In section \ref{o2}, we obtain the cosmological parameters like the  tensor to scalar ratio, and compare with experimental data.

\section{ Emergence of $F(R)$-gravity on an (11+3)-dimensional manifold}\label{o1}

In this section, we will assume that our universe has been constructed on a D3-brane. Thus, the action of gravity and matter in a 4-dimensional universe should be equal to the action of a D3-brane (See Figure 1). This means that strings which live on a D3-brane produces gravity and different types of matter. Thus, each string ($X^{i}$) can be expanded in terms of curvatures ($R$), gauge fields ($F$) and scalars ($\phi$) (See Figure 2). According to the Horava-Witten mechanism \cite{b1,b2}, this D3-brane can be a part of a bigger 11-dimensional manifold on which the action of fields should be anomaly free (See Figure 3). However, using the relation between strings and fields, we notice that the gauge variation of the action  on this manifold is not zero, and some extra anomalies  emerge. To remove these anomalies, we have to add an extra 11-dimensional manifold which is connected with the first 11-dimensional manifold by a 3-dimensional manifold. The extra fields which are produced by the interaction of the two manifolds produce  $F(R)$-gravity (See Figure 4).

\begin{figure*}[thbp]
	\begin{center}
		\begin{tabular}{rl}
			\includegraphics[width=8cm]{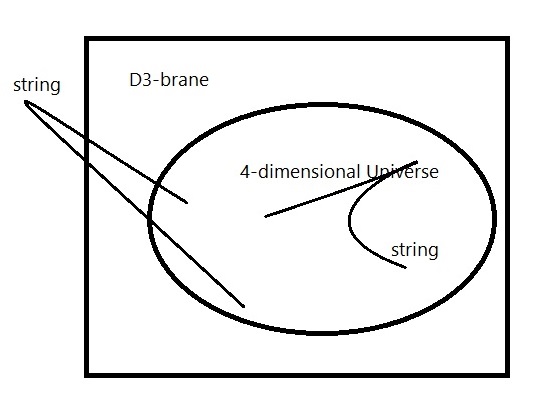}
		\end{tabular}
	\end{center}
	\caption{  4-dimensional universe on D3-brane. }
\end{figure*}

\begin{figure*}[thbp]
	\begin{center}
		\begin{tabular}{rl}
			\includegraphics[width=8cm]{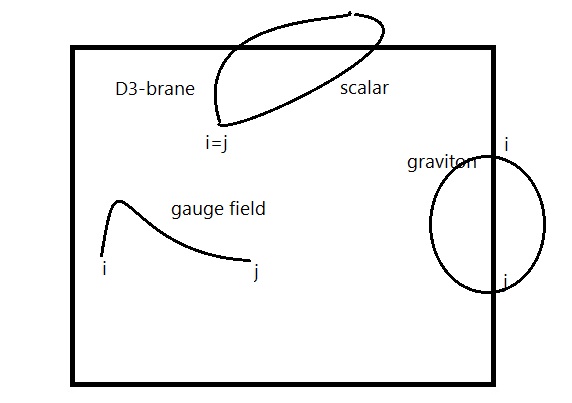}
		\end{tabular}
	\end{center}
	\caption{  Different shapes of strings produce different types of fields. }
\end{figure*}

\begin{figure*}[thbp]
	\begin{center}
		\begin{tabular}{rl}
			\includegraphics[width=8cm]{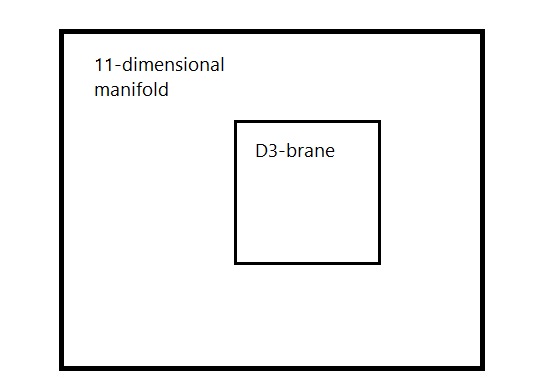}
		\end{tabular}
	\end{center}
	\caption{  D3-brane is a part of 11-dimensional manifold. }
\end{figure*}

\begin{figure*}[thbp]
	\begin{center}
		\begin{tabular}{rl}
			\includegraphics[width=8cm]{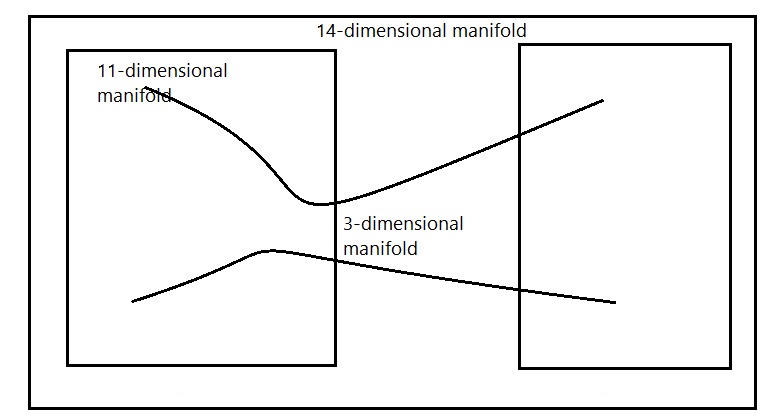}
		\end{tabular}
	\end{center}
	\caption{  Two 11-dimensional manifolds interact via 3-dimensional manifold in 14-dimensional manifold. }
\end{figure*}

First, let us introduce the action of the D3-brane which is given by \cite{D3}:

\begin{eqnarray}
S_{D3}&=& -T_{D3}\int d^{4}y \sqrt{-det(\bar{\gamma}_{ab}+2\pi l_{s}^{2}F_{ab})}, \nonumber\\
\bar{\gamma}_{ab}&=&g_{\mu\nu}\partial_{a}X^{\mu}\partial_{b}X^{\nu}, \nonumber\\ F_{ab} &=& \partial_{a}A_{b}-\partial_{b}A_{a} \label{D1}
\end{eqnarray}
where $A_{b}$ is the gauge field, $F_{ab}$ is the field strength, $X^{\mu}$ is the string, $g_{\mu\nu}$ is the metric, $T_{D3}$ is the tension and $l_{s}$ is the string length. Substituting $X^{0}=t$ and doing some mathematical calculations, the acion of the D3-brane in equation (\ref{D1}) is given by \cite{D3}:

\begin{eqnarray}
S_{D3}&=& -T_{D3}\int d^{4}y \sqrt{1+ g_{ij}\partial_{a}X^{i}\partial^{a}X^{j} -4\pi^{2} l_{s}^{4}F_{ab}F^{ab}},  \label{D2}
\end{eqnarray}

To construct our universe on a D3-brane, this action should be equal to the action of fields and gravity in a 4-dimensional universe. The action of matter and gravity is given by:

\begin{eqnarray}
S_{Gravity-Matter}&=&- \int d^{4}y \sqrt{-g}\Big(R -g_{ab}\partial^{a}\phi\partial^{b}\phi -i\bar{\psi}\gamma^{a}\partial_{a}\psi+1\Big),  \label{D3}
\end{eqnarray}
where $\phi$ is the scalar field and $\psi$ is the fermionic field. Puting equation (\ref{D2}) equal to equation (\ref{D3}), and doing some mathematical calculations, we obtain the relation between strings and matter fields:

\begin{eqnarray}
S_{Gravity-Matter}&=& S_{D3}  \nonumber\\ \Longrightarrow X^{i} &=& \int dy^{i} \sqrt{-1 + 4\pi^{2} l_{s}^{4}F_{ij}F^{ij} + \Big(\sqrt{-g}(g^{ij}R_{ij} -g_{ij}\partial^{i}\phi\partial^{j}\phi -i\bar{\psi}\gamma^{i}\partial_{i}\psi+1)\Big)^{2}},  \label{D4}
\end{eqnarray}
where we have assumed $T_{D3}\simeq 1$ and $4\pi^{2} l_{s}^{4}\simeq 1$. Using the above relation between matter fields and strings, we can reconsider the Horava-Witten mechanism. We will show that there are also other anomalies that have been ignored in previous considerations.  Our goal is to show that by adding a 3-dimensional manifold to 11-dimensional spacetime in the Horava-Witten mechanism, all anomalies can be removed and an action without anomaly can be produced. This action is identical to the action of the modified gravity  theory presented in \cite{p1,p2,p3,p4,p5,p6}.

At this stage, we introduce the Horava-Witten mechanism in  11-dimensional spacetime. In this theory, the bosonic part of the action in 11-dimensional supergravity (SUGRA) is given by \cite{b1,b2}:

     \begin{eqnarray}
     S_{\rm Bosonic-SUGRA}&=& \frac{1}{\bar{\kappa}^{2}}\int d^{11}x\sqrt{g}\Big(-\frac{1}{2}R-\frac{1}{48}G_{IJKL}G^{IJKL}\Big) + S_{\rm CGG}, \nonumber\\
     S_{\rm CGG}&=&-\frac{\sqrt{2}}{3456\bar{\kappa}^{2}}\int_{M^{11}}d^{11}x \varepsilon^{I_{1}I_{2}...I_{11}}C_{I_{1}I_{2}I_{3}}G_{I_{4}...I_{7}}G_{I_{8}...I_{11}}, \label{s1}
     \end{eqnarray}
where $\varepsilon^{I_1I_2..I_k}$ is the rank-$k$ Levi-Civita pseudotensor,  and CGG is used to denote the product term of the three-form field $C_{I_{1}I_{2}I_{3}}$ and four-form field $G_{IJKL}$, which are directly related to the gauge field $A^I$, field strength $F^{IJ}$ and Ricci curvature $R^{IJ}$ \cite{b2}:

     \begin{eqnarray}
      G_{IJKL}&=&-\frac{3}{\sqrt{2}}\frac{\kappa^{2}}{\lambda^{2}}\varepsilon(x^{11})(F_{[IJ}F_{KL]}-R_{[IJ}R_{KL]})+...,\nonumber\\
      \delta C_{ABC}&=&-\frac{\kappa^{2}}{6\sqrt{2}\lambda^{2}}\delta (x^{11})\tr(\epsilon_C F_{AB}-\epsilon_C R_{AB}),\nonumber\\
      G_{11ABC}&=&(\partial_{11}C_{ABC}\pm \text{23 permutations of the indices $11$ and $ABC$})+\frac{\kappa^{2}}{\sqrt{2}\lambda^{2}}\delta (x^{11})\omega_{ABC},\nonumber\\
      \delta \omega_{ABC}&=&\partial_{A}\tr(\epsilon F_{BC})+ \text{cyclic permutations of}~ABC,\nonumber\\
      F^{IJ}&=&\partial^{I}A^{J}-\partial^{J}A^{I},\nonumber\\
      R_{IJ}&=&\partial_{I}\Gamma^{B}_{JB}-\partial_{J}\Gamma^{B}_{IB} +\Gamma^{A}_{JB}\Gamma^{B}_{IA} -\Gamma^{A}_{IB}\Gamma^{B}_{JA},\nonumber\\
      \Gamma_{IJK}&=&\partial_{I}g_{JK}+\partial_{K}g_{IJ}-\partial_{J}g_{IK}, \nonumber\\
      G_{IJ}&=&R_{IJ}-\frac{1}{2}R g_{IJ},\label{s2}
      \end{eqnarray}
where $\epsilon$ and $\epsilon_C$ characterize infinitesimal gauge transformations \cite{b2}. Here,  $\varepsilon(x^{11})$ is 1 for $x^{11}> 0$ and $-1$ for $x^{11}< 0$ and also $\delta(x^{11})=\frac{1}{2}\partial \varepsilon(x^{11})/\partial x^{11}$ is the Dirac delta function. As usual \cite{b2}, $\tr$ is 1/30th of the trace $\Tr$ in the adjoint representation for $E_8\times E_8$. The ellipsis ($...$) denotes terms that are regular near $x^{11}=0$ and hence vanish there \cite{b2}. It is clear from the above equation that G-fields can be produced by joining F-fields or R-fields. This means that G-fields are produced by joining two strings,  each of which produces one gauge field or curvature ($R$) (See Figure 5).

\begin{figure*}[thbp]
	\begin{center}
		\begin{tabular}{rl}
			\includegraphics[width=8cm]{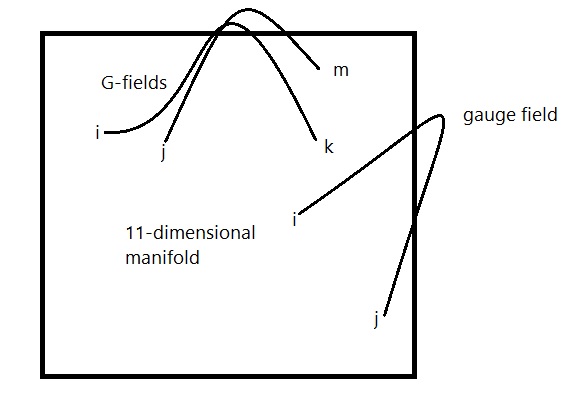}
		\end{tabular}
	\end{center}
	\caption{  G-fields are produced by joining two strings. }
\end{figure*}

The gauge variation of the 
CGG term in the action yields the following equation \cite{b2}:

      \begin{eqnarray}
        \delta S_{\rm CGG}|_{11}&=&-\frac{\sqrt{2}}{3456\bar{\kappa}^{2}}\int_{M^{11}}d^{11}x \varepsilon^{I_{1}I_{2}...I_{11}}\delta C_{I_{1}I_{2}I_{3}}G_{I_{4}...I_{7}}G_{I_{8}...I_{11}}\nonumber\\
        &\approx & - \frac{\bar{\kappa}^{4}}{128 \lambda^{6}}\int_{M^{10}}\Sigma_{n=1}^{5}(\tr F^{n}-\tr R^{n}+\tr(F^{n}R^{5-n})),\label{s3}
        \end{eqnarray}
%
      where $\tr F^{n}=\tr(F_{[I_{1}I_{2}}..F_{I_{2n-1}I_{2n}]})=\varepsilon^{I_{1}I_{2}..I_{2n-1}I_{2n}}F_{I_{1}I_{2}}..F_{I_{2n-1}I_{2n}}$ or $\tr R^{n}=\tr(R_{[I_{1}I_{2}}..R_{I_{2n-1}I_{2n}]})=\varepsilon^{I_{1}I_{2}..I_{2n-1}I_{2n}}R_{I_{1}I_{2}}..R_{I_{2n-1}I_{2n}}$ . The above terms cancel the  anomaly of  ($S_{\rm Bosonic-SUGRA}$) in 11-dimensional manifolds \cite{b2}:

      \begin{eqnarray}
          && \delta S_{\rm CGG}|_{11}=
          -\delta S^{\rm anomaly}_{\rm Bosonic-SUGRA}.\label{ss3}
          \end{eqnarray}

       Thus, $S_{\rm CGG}$ is necessary for anomaly cancellation. Our goal now is to find a good rationale for its inclusion. We also answer  the issue of the relation between  CGG terms in 11-dimensional supergravity and $F(R)$-gravity. In fact, we suggest a theory in which $F(R)$-gravity terms appear in the supergravity action without being added by hand. To this end, we  choose a unified form for all particles and fields by using Nambu-Poisson brackets and properties of string fields ($X$). Solving equation (\ref{D4}) and assuming that changes of fields with respect to the coordinates be ingnorable, we obtain \cite{b9,A1}:

      \begin{eqnarray}       
              X^{I}&\approx&y^{I}+ y_{J}F^{IJ} - y_{J}R^{IJ} + y_{J}\partial^{I}\phi\partial^{J}\phi -iy_{J} \bar{\psi}\gamma^{J}\partial^{I}\psi  +...\nonumber\\
      \Rightarrow \{ X^{I},X^{J}\}&=&\Sigma_{I,J}\varepsilon^{I'J'}\frac{\partial X^{I}}{\partial y^{I'}}\frac{\partial X^{J}}{\partial y^{J'}}\nonumber\\
       &=&F^{IJ} - R^{IJ} + \partial^{I}\phi\partial^{J}\phi -i \bar{\psi}\gamma^{J}\partial^{I}\psi  +....\label{s4}
       \end{eqnarray}
where we introduced the scalar field $\phi$. Also,  $A^{I}$ is the gauge field and $R$ is the curvature (R). In fact, the origin of all matter  is the same and they are different shapes of strings.  In the static state, all strings can be described by a unit vector ($X^{i}$). When these strings interact with each other or move, the initial symmetry is broken and fields and particles emerge. Using 4-dimensional instead of 2-dimensional brackets, we may derive the GG term $G_{IJKL}G^{IJKL}$ in supergravity in terms of strings ($X$) \cite{b9,A1}:

          \begin{eqnarray}
          G^{IJKL}&=& \{ X^{I},X^{J},X^{K},X^{L} \} =
          \varepsilon^{I'J'K'L'}\frac{\partial X^{I}}{\partial y^{I'}}\frac{\partial X^{J}}{\partial y^{J'}}\frac{\partial X^{K}}{\partial y^{K'}}\frac{\partial X^{L}}{\partial y^{L'}}\nonumber\\
          &\Downarrow&\nonumber\\
          \int d^{11}x\sqrt{g}\Big(G_{IJKL}G^{IJKL}\Big)&=& 
          \int d^{11}x\sqrt{g}\Big(
          \varepsilon_{I'J'K'L'}\frac{\partial X_{I}}{\partial y_{I'}}\frac{\partial X_{J}}{\partial y_{J'}}\frac{\partial X_{K}}{\partial y_{K'}}\frac{\partial X_{L}}{\partial y_{L'}}
          \varepsilon^{I''J''K''L''}\frac{\partial X^{I}}{\partial y^{I''}}\frac{\partial X^{J}}{\partial y^{J''}}\frac{\partial X^{K}}{\partial y^{K''}}\frac{\partial X^{L}}{\partial y^{L''}}\Big).\label{s13}
          \end{eqnarray}
       Equation (\ref{s13}) helps us to extract the CGG terms from the GG terms in supergravity. To this aim, we will add a 3-dimensional manifold   to the 11-dimensional manifold which connects it to other 11-dimensional manifold by applying the properties of strings ($X$) in the Nambu-Poisson brackets \cite{A1}:

           \begin{eqnarray}
          X^{I}&\approx&y^{I}+ y_{J}F^{IJ} - y_{J}R^{IJ} + y_{J}\partial^{I}\phi\partial^{J}\phi -iy_{J} \bar{\psi}\gamma^{J}\partial^{I}\psi  +...\nonumber\\
           &\Downarrow&\nonumber\\
           \frac{\partial X^{I_{5}}}{\partial y^{I_{5}}}&\approx&\delta ( y^{I_{5}})+... \quad \frac{\partial X^{I_{6}}}{\partial y^{I_{6}}}\approx\delta ( y^{I_{6}})+... \quad \frac{\partial X^{I_{7}}}{\partial y^{I_{7}}}\approx\delta ( y^{I_{7}})+...,\\
           \int_{M^{N=3}}&\rightarrow&\int_{y^{I_{5}}+y^{I_{6}}+y^{I_{7}}}\varepsilon^{I'_{5}I'_{6}I'_{7}}\frac{\partial X^{I_{5}}}{\partial y^{I'_{5}}}\frac{\partial X^{I_{6}}}{\partial y^{I'_{6}}}\frac{\partial X^{I_{7}}}{\partial y^{I'_{7}}}=1+..., \label{s14}
           \end{eqnarray}
where ellipses (...) were used to represent higher-order derivatives. The integration is over a 3-dimensional manifold with coordinates ($y^{I_{5}},y^{I_{6}},y^{I_{7}}$) and consequently, the integration can be shown by $\int_{y^{I_{5}}+y^{I_{6}}+y^{I_{7}}}=\int dy^{I_{5}}\int dy^{I_{6}}\int dy^{I_{7}}$). This result shows that ignoring fluctuations of strings that leads to the production of fields, the result of integration over each 3-dimensional manifold tends to one. When we add one manifold to another, the integration will be the product of integration over each manifold.

Extending the manifold over additional dimensions extends the integration volume element.  By extending the 11-dimensional manifold in Eq.~(\ref{s13}) with the 3-dimensional manifold of Eq.~(\ref{s14}), we get \cite{A1}:

            \begin{eqnarray}
            && \int_{M^{N=3}} \int_{M^{11}}\sqrt{g}\Big(G_{I_{1}I_{2}I_{3}I_{4}}G^{I_{1}I_{2}I_{3}I_{4}}\Big)
             =\int_{M^{11}+y^{I_{5}}+y^{I_{6}}+y^{I_{7}}}\sqrt{g}\epsilon^{I'_{5}I'_{6}I'_{7}} G_{I_{1}I_{2}I_{3}I_{4}}G^{I_{1}I_{2}I_{3}I_{4}}\frac{\partial X^{I_{5}}}{\partial y^{I'_{5}}}\frac{\partial X^{I_{6}}}{\partial y^{I'_{6}}}\frac{\partial X^{I_{7}}}{\partial y^{I'_{7}}},
\end{eqnarray}
where we notice that after making the identification \cite{A1}:
            \begin{eqnarray}
            C^{I_{5}I_{6}I_{7}}&=& 
            \epsilon^{I'_{5}I'_{6}I'_{7}}\frac{\partial X^{I_{5}}}{\partial y^{I'_{5}}}\frac{\partial X^{I_{6}}}{\partial y^{I'_{6}}}\frac{\partial X^{I_{7}}}{\partial y^{I'_{7}}},\label{s15}
            \end{eqnarray}
we recover the CGG action in $(11+3)$ dimensions \cite{A1}:
\begin{equation}
S^{N=11+3}_{\rm CGG}=\int_{M^{11}+y^{I_{5}}+y^{I_{6}}+y^{I_{7}}}\sqrt{g} G_{I_{1}I_{2}I_{3}I_{4}}G^{I_{1}I_{2}I_{3}I_{4}}
C^{I_{5}I_{6}I_{7}}\label{st15}.
\end{equation}

This equation has three interesting results:
\begin{inparaenum}[1.]
\item The CGG term that appears in the supergravity action is  a result of adding a 3-dimensional manifold to  an 11-dimensional manifold. This manifold connects the first 11-dimensional manifold to the second one. \item Combining the 11-dimensional manifold with the 3-dimensional manifold yields 14-dimensional supergravity.
\item The shape of the C-term is now clear in terms of string fields ($X^{I}$).
\item It is clear that C-fields are in fact G-fields such that one end of the  strings is placed out of the 11-dimensional manifold and on the 3-dimensional manifold and for this reason, C-fields seem to have three ends only (See Figure 6). 
\end{inparaenum}

\begin{figure*}[thbp]
	\begin{center}
		\begin{tabular}{rl}
			\includegraphics[width=8cm]{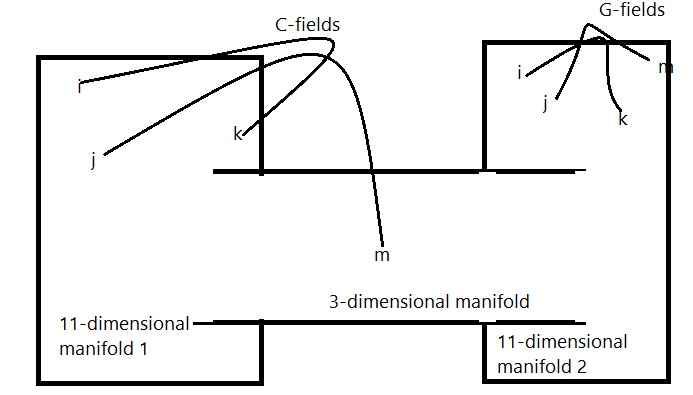}
		\end{tabular}
	\end{center}
	\caption{  C-fields are produced from G-fields when one end of strings is placed on a 3-dimensional manifold and three ends are placed on an  11-dimensional manifold. }
\end{figure*}

       To verify that  the theory is correct, we should be able to get back the gauge variation of the CGG-action in Eq.~(\ref{s3}) in terms of field  strengths and curvature. To this end, using Eqs.~(\ref{s4}, \ref{s13}, \ref{s14} and \ref{s15}), we can calculate the gauge variation of C \cite{A1}:

             \begin{eqnarray}
           X^{I}&\approx&y^{I}+ y_{J}F^{IJ} - y_{J}R^{IJ} + y_{J}\partial^{I}\phi\partial^{J}\phi -iy_{J} \bar{\psi}\gamma^{J}\partial^{I}\psi  +...\nonumber\\
             &\Downarrow&\nonumber\\
             \frac{\partial \delta_{A} X^{I}}{\partial y^{I}}&=&\delta ( y^{I}) \nonumber\\
             &\Downarrow&\nonumber\\
             \int_{M^{N=3}+M^{11}}\delta_{A} C^{I_{5}I_{6}I_{7}}&=&\int_{M^{N=3}+M^{11}} \Sigma_{I'_{5}I'_{6}I'_{7}}\varepsilon^{I'_{5}I'_{6}I'_{7}}\delta_{A}(\frac{\partial X^{I_{5}}}{\partial y^{I'_{5}}}\frac{\partial X^{I_{6}}}{\partial y^{I'_{6}}}\frac{\partial X^{I_{7}}}{\partial y^{I'_{7}}})\nonumber\\
             &=& \int_{M^{N=3}+M^{10}}\Sigma_{I'_{5}I'_{6}}\varepsilon^{I'_{5}I'_{6}}(\frac{\partial X^{I_{5}}}{\partial y^{I'_{5}}}\frac{\partial X^{I_{6}}}{\partial y^{I'_{6}}})\nonumber\\
             &=&\int_{M^{N=3}+M^{10}}(F^{IJ} - R^{IJ} + \partial^{I}\phi\partial^{J}\phi -i \bar{\psi}\gamma^{J}\partial^{I}\psi  +....\nonumber\\
             &=&\int_{M^{N=3}+M^{10}}(F^{IJ} - R^{IJ} + \partial^{I}\phi\partial^{J}\phi -i \bar{\psi}\gamma^{J}\partial^{I}\psi  +....).\label{s16}
             \end{eqnarray}
where the ellipses (...) represent higher-order derivatives with respect to fields. 
Using Eqs.~(\ref{s4}, \ref{s13},  \ref{s14}, \ref{s15} and \ref{s16})  we can calculate the gauge variation of the CGG action in the equation of (\ref{st15}):
\begin{eqnarray}
\delta S^{N=11+3}_{\rm CGG}&=&
\delta\int_{M^{11}+M^{N=3}}\sqrt{g}\epsilon_{I_{1}I_{2}I_{3}I_{4}I'_{1}I'_{2}I'_{3}I'_{4}I_{5}I_{6}I_{7}}\epsilon^{\tilde{I}_{5}\tilde{I}_{6}\tilde{I}_{7}}(\frac{\partial X^{I_{5}}}{\partial y^{\tilde{I}_{5}}}\frac{\partial X^{I_{6}}}{\partial y^{\tilde{I}_{6}}}\frac{\partial X^{I_{7}}}{\partial y^{\tilde{I}_{7}}}) G^{I_{1}I_{2}I_{3}I_{4}}G^{I'_{1}I'_{2}I'_{3}I'_{4}}\nonumber\\
&=&\delta\int_{M^{11}+M^{N=3}}\sqrt{g}\epsilon_{I_{1}I_{2}I_{3}I_{4}I'_{1}I'_{2}I'_{3}I'_{4}I_{5}I_{6}I_{7}}\epsilon^{\tilde{I}_{5}\tilde{I}_{6}\tilde{I}_{7}}(\frac{\partial X^{I_{5}}}{\partial y^{\tilde{I}_{5}}}\frac{\partial X^{I_{6}}}{\partial y^{\tilde{I}_{6}}}\frac{\partial X^{I_{7}}}{\partial y^{\tilde{I}_{7}}})\nonumber\\
&&\hskip 0.5in{}\times(\epsilon^{\tilde{I}_{1}\tilde{I}_{2}\tilde{I}_{3}\tilde{I}_{4}}\frac{\partial X^{I_{1}}}{\partial y^{\tilde{I}_{1}}}\frac{\partial X^{I_{2}}}{\partial y^{\tilde{I}_{2}}}\frac{\partial X^{I_{3}}}{\partial y^{\tilde{I}_{3}}}\frac{\partial X^{I_{4}}}{\partial y^{\tilde{I}_{4}}})(\epsilon^{\tilde{I}'_{1}\tilde{I}'_{2}\tilde{I}'_{3}\tilde{I}'_{4}}\frac{\partial X^{I'_{1}}}{\partial y^{\tilde{I}'_{1}}}\frac{\partial X^{I'_{2}}}{\partial y^{\tilde{I}'_{2}}}\frac{\partial X^{I'_{3}}}{\partial y^{\tilde{I}'_{3}}}\frac{\partial X^{I'_{4}}}{\partial y^{\tilde{I}'_{4}}})\nonumber\\
&=&\int_{M^{10}+M^{N=3}}\sqrt{g}\epsilon_{I_{1}I_{2}I_{3}I_{4}I'_{1}I'_{2}I'_{3}I'_{4}I_{5}I_{6}}\epsilon^{\tilde{I}_{5}\tilde{I}_{6}}(\frac{\partial X^{I_{5}}}{\partial y^{\tilde{I}_{5}}}\frac{\partial X^{I_{6}}}{\partial y^{\tilde{I}_{6}}})\nonumber\\
&&\hskip 0.5in{}\times(\epsilon^{\tilde{I}_{1}\tilde{I}_{2}\tilde{I}_{3}\tilde{I}_{4}}\frac{\partial X^{I_{1}}}{\partial y^{\tilde{I}_{1}}}\frac{\partial X^{I_{2}}}{\partial y^{\tilde{I}_{2}}}\frac{\partial X^{I_{3}}}{\partial y^{\tilde{I}_{3}}}\frac{\partial X^{I_{4}}}{\partial y^{\tilde{I}_{4}}})(\epsilon^{\tilde{I}'_{1}\tilde{I}'_{2}\tilde{I}'_{3}\tilde{I}'_{4}}\frac{\partial X^{I'_{1}}}{\partial y^{\tilde{I}'_{1}}}\frac{\partial X^{I'_{2}}}{\partial y^{\tilde{I}'_{2}}}\frac{\partial X^{I'_{3}}}{\partial y^{\tilde{I}'_{3}}}\frac{\partial X^{I'_{4}}}{\partial y^{\tilde{I}'_{4}}})\nonumber\\
&=&\int_{M^{10}+M^{N=3}}\sqrt{g}\epsilon_{I_{1}I_{2}I_{3}I_{4}I'_{1}I'_{2}I'_{3}I'_{4}I_{5}I_{6}}(\epsilon^{\tilde{I}_{4}\tilde{I}_{5}}\frac{\partial X^{I_{4}}}{\partial y^{\tilde{I}_{4}}}\frac{\partial X^{I_{5}}}{\partial y^{\tilde{I}_{5}}})(\epsilon^{\tilde{I}'_{4}\tilde{I}_{6}}\frac{\partial X^{I'_{4}}}{\partial y^{\tilde{I}'_{4}}}\frac{\partial X^{I_{6}}}{\partial y^{\tilde{I}_{6}}})\nonumber\\
&&\hskip 0.5in{}\times(\epsilon^{\tilde{I}_{1}\tilde{I}_{2}}\frac{\partial X^{I_{1}}}{\partial y^{\tilde{I}_{1}}}\frac{\partial X^{I_{2}}}{\partial y^{\tilde{I}_{2}}})(\epsilon^{\tilde{I}'_{1}\tilde{I}'_{2}}\frac{\partial X^{I'_{1}}}{\partial y^{\tilde{I}'_{1}}}\frac{\partial X^{I'_{2}}}{\partial y^{\tilde{I}'_{2}}})(\epsilon^{\tilde{I}_{3}\tilde{I}'_{3}}\frac{\partial X^{I_{3}}}{\partial \tilde{I}^{I_{3}}}\frac{\partial X^{I'_{3}}}{\partial y^{\tilde{I}'_{3}}})\nonumber\\
&=&\int_{M^{10}+M^{N=3}}\sqrt{g}\Sigma_{n=1}^{5}\Big(\tr F^{n}-\tr R^{n}+\tr F^{n}R^{5-n}\Big)\nonumber\\
 &+&\int_{M^{11}+M^{N=3}}\sqrt{g}\Big(\Sigma_{n=1}^{5}\Sigma_{m=0}^{5-n} \varepsilon^{J_{1}J_{2}..J_{2n-1}J_{2n}}R_{J_{1}J_{2}}..R_{J_{2n-1}J_{2n}}\varepsilon^{I_{1}I_{2}..I_{2m-1}I_{2m}}\partial^{I_{1}}\phi\partial^{I_{2}}\phi..\partial^{I_{2m-1}}\phi\partial^{I_{2m}}\phi\Big) \nonumber\\
 &-&i^{m}\int_{M^{11}+M^{N=3}}\sqrt{g}\Big(\Sigma_{n=1}^{5}\Sigma_{m=0}^{5-n} \varepsilon^{J_{1}J_{2}..J_{2n-1}J_{2n}}R_{J_{1}J_{2}}..R_{J_{2n-1}J_{2n}}\varepsilon^{I_{1}I_{2}..I_{2m-1}I_{2m}}\bar{\psi} \gamma^{I_{1}}\partial^{I_{2}}\psi..\bar{\psi} \gamma^{2I_{m}-1}\partial^{I_{2m}}\psi\Big)\nonumber\\
 &-&i^{m}\int_{M^{11}+M^{N=3}}\sqrt{g}\Big(\Sigma_{n=1}^{5}\Sigma_{m=0}^{n-5}\Sigma_{k=0}^{5-n-m} \varepsilon^{J_{1}J_{2}..J_{2n-1}J_{2n}}R_{J_{1}J_{2}}..R_{J_{2n-1}J_{2n}}\varepsilon^{I_{1}I_{2}..I_{2m-1}I_{2m}}\bar{\psi} \gamma^{I_{1}}\partial^{I_{2}}\psi..\bar{\psi} \gamma^{2I_{m}-1}\partial^{I_{2m}}\psi  \nonumber\\
&\times& \varepsilon^{I_{1}I_{2}..I_{2k-1}I_{2k}}\partial^{I_{1}}\phi\partial^{I_{2}}\phi..\partial^{I_{2k-1}}\phi\partial^{I_{2k}}\phi \Big)\nonumber\\
&+&\int_{M^{11}+M^{N=3}}\sqrt{g}\Big(\Sigma_{n=1}^{5}\Sigma_{m=0}^{5-n} \varepsilon^{J_{1}J_{2}..J_{2n-1}J_{2n}}F_{J_{1}J_{2}}..F_{J_{2n-1}J_{2n}}\varepsilon^{I_{1}I_{2}..I_{2m-1}I_{2m}}\partial^{I_{1}}\phi\partial^{I_{2}}\phi..\partial^{I_{2m-1}}\phi\partial^{I_{2m}}\phi \Big)\nonumber\\
&-&i^{m}\int_{M^{11}+M^{N=3}}\sqrt{g}\Big(\Sigma_{n=1}^{5}\Sigma_{m=0}^{5-n} \varepsilon^{J_{1}J_{2}..J_{2n-1}J_{2n}}F_{J_{1}J_{2}}..F_{J_{2n-1}J_{2n}}\varepsilon^{I_{1}I_{2}..I_{2m-1}I_{2m}}\bar{\psi} \gamma^{I_{1}}\partial^{I_{2}}\psi..\bar{\psi} \gamma^{2I_{m}-1}\partial^{I_{2m}}\psi \Big)\nonumber\\
&-&i^{m}\int_{M^{11}+M^{N=3}}\sqrt{g}\Big(\Sigma_{n=1}^{5}\Sigma_{m=0}^{5-n}\Sigma_{k=0}^{5-n-m} \varepsilon^{J_{1}J_{2}..J_{2n-1}J_{2n}}F_{J_{1}J_{2}}..F_{J_{2n-1}J_{2n}}\varepsilon^{I_{1}I_{2}..I_{2m-1}I_{2m}}\bar{\psi} \gamma^{I_{1}}\partial^{I_{2}}\psi..\bar{\psi} \gamma^{2I_{m}-1}\partial^{I_{2m}}\psi  \nonumber\\
&\times& \varepsilon^{I_{1}I_{2}..I_{2k-1}I_{2k}}\partial^{I_{1}}\phi\partial^{I_{2}}\phi..\partial^{I_{2k-1}}\phi\partial^{I_{2k}}\phi+...\Big).\label{s17}
\end{eqnarray}
The first term in Eq.~(\ref{s17}) cancels the anomaly in Eq.~(\ref{s3}). However, the other terms yield the action of modified gravity and matter. In fact, to remove the anomalies, we have to add two extra actions to equation (\ref{s1}), one is related to $F(R)$-gravity, and the second is corresponds to matter. We can write:

 \begin{eqnarray}
S_{\rm SUGRA}&=&  S_{\rm GG} + S_{\rm CGG} + S_{\rm F(R)} + S_{\rm Matter},\nonumber\\ S_{\rm GG}&=&\frac{1}{\bar{\kappa}^{2}}\int d^{14}x\sqrt{g}\Big(-\frac{1}{2}R-\frac{1}{48}G_{IJKL}G^{IJKL}\Big),  \nonumber\\
S_{\rm CGG}&=&-\frac{\sqrt{2}}{3456\bar{\kappa}^{2}}\int_{M^{11}}d^{14}x \varepsilon^{I_{1}I_{2}...I_{11}}C_{I_{1}I_{2}I_{3}}G_{I_{4}...I_{7}}G_{I_{8}...I_{11}}, \nonumber\\
S_{\rm F(R)}&=& \int_{M^{11}+M^{N=3}}\sqrt{g}F(R,\phi,\psi)\nonumber\\
S_{\rm Matter}&=& \int_{M^{11}+M^{N=3}}\sqrt{g}\Big(\Sigma_{n=1}^{6}\Sigma_{m=0}^{n-6} \varepsilon^{J_{1}J_{2}..J_{2n-1}J_{2n}}F_{J_{1}J_{2}}..F_{J_{2n-1}J_{2n}}\varepsilon^{I_{1}I_{2}..I_{2m-1}I_{2m}}\partial^{I_{1}}\phi\partial^{I_{2}}\phi..\partial^{I_{2m-1}}\phi\partial^{I_{2m}}\phi \Big)\nonumber\\
&-&i^{m}\int_{M^{11}+M^{N=3}}\sqrt{g}\Big(\Sigma_{n=1}^{6}\Sigma_{m=0}^{6-n} \varepsilon^{J_{1}J_{2}..J_{2n-1}J_{2n}}F_{J_{1}J_{2}}..F_{J_{2n-1}J_{2n}}\varepsilon^{I_{1}I_{2}..I_{2m-1}I_{2m}}\bar{\psi} \gamma^{I_{1}}\partial^{I_{2}}\psi..\bar{\psi} \gamma^{2I_{m}-1}\partial^{I_{2m}}\psi \Big)\nonumber\\
&-&i^{m}\int_{M^{11}+M^{N=3}}\sqrt{g}\Big(\Sigma_{n=1}^{6}\Sigma_{m=0}^{6-n}\Sigma_{k=0}^{6-n-m} \varepsilon^{J_{1}J_{2}..J_{2n-1}J_{2n}}F_{J_{1}J_{2}}..F_{J_{2n-1}J_{2n}}\varepsilon^{I_{1}I_{2}..I_{2m-1}I_{2m}}\bar{\psi} \gamma^{I_{1}}\partial^{I_{2}}\psi..\bar{\psi} \gamma^{2I_{m}-1}\partial^{I_{2m}}\psi  \nonumber\\
&\times& \varepsilon^{I_{1}I_{2}..I_{2k-1}I_{2k}}\partial^{I_{1}}\phi\partial^{I_{2}}\phi..\partial^{I_{2k-1}}\phi\partial^{I_{2k}}\phi+...\Big)\label{s18}
\end{eqnarray}
where $F(R,\phi,\psi)$ can be given by :

 \begin{eqnarray}
F(R,\phi,\psi) &=&\Big(\Sigma_{n=1}^{6}\Sigma_{m=0}^{n-6} \varepsilon^{J_{1}J_{2}..J_{2n-1}J_{2n}}R_{J_{1}J_{2}}..R_{J_{2n-1}J_{2n}}\varepsilon^{I_{1}I_{2}..I_{2m-1}I_{2m}}\partial^{I_{1}}\phi\partial^{I_{2}}\phi..\partial^{I_{2m-1}}\phi\partial^{I_{2m}}\phi \Big)\nonumber\\
&-&i^{m}\Big(\Sigma_{n=1}^{6}\Sigma_{m=0}^{6-n} \varepsilon^{J_{1}J_{2}..J_{2n-1}J_{2n}}R_{J_{1}J_{2}}..R_{J_{2n-1}J_{2n}}\varepsilon^{I_{1}I_{2}..I_{2m-1}I_{2m}}\bar{\psi} \gamma^{I_{1}}\partial^{I_{2}}\psi..\bar{\psi} \gamma^{2I_{m}-1}\partial^{I_{2m}}\psi \Big)\nonumber\\
&-&i^{m}\Big(\Sigma_{n=1}^{6}\Sigma_{m=0}^{6-n}\Sigma_{k=0}^{6-n-m} \varepsilon^{J_{1}J_{2}..J_{2n-1}J_{2n}}R_{J_{1}J_{2}}..R_{J_{2n-1}J_{2n}}\varepsilon^{I_{1}I_{2}..I_{2m-1}I_{2m}}\bar{\psi} \gamma^{I_{1}}\partial^{I_{2}}\psi..\bar{\psi} \gamma^{2I_{m}-1}\partial^{I_{2m}}\psi  \nonumber\\
&\times& \varepsilon^{I_{1}I_{2}..I_{2k-1}I_{2k}}\partial^{I_{1}}\phi\partial^{I_{2}}\phi..\partial^{I_{2k-1}}\phi\partial^{I_{2k}}\phi+...\Big)\label{s19}
\end{eqnarray}

This $F(R,\phi,\psi)$-gravity is now free of any anomaly on the (11 + 3)-dimensional manifold. In fact, this type of gravity is produced by exchanging gravitons, scalars and fermions between two 11-dimensional manifolds via one 3-dimensional manifold. The order of curvatures depends on the number of dimensions of the manifolds and also on the  number of scalars or fermions which are attached to the curvatures. This result is in good agreement with previous predictions in \cite{p7,p8}.

   \section{Inflationary universe in $F(R,\phi,\psi)$-gravity }\label{o2}
   
   In this section, we construct a  4-dimensional universe on a (11+3)-dimensional manifold. We will show that interaction between two 11-dimensional manifolds in this system has a direct effect on the evolution of the universe during the inflationary epoch. We will obtain the scale factor of the universe in terms of the parameters of the system. Using the action in equation of (\ref{s18}) and the relation between $G$, $C$, $F$ and $R$-terms in equation (\ref{s2}), we can obtain the following field equations:

\begin{eqnarray}
&&R_{ij}-\frac{1}{2} R g_{ij} \nonumber\\
&+& (\frac{1}{2}  g_{ij}F(R,\phi,\psi) - R_{ij}F'(R,\phi,\psi) + \partial_{i}\partial_{j}F'(R,\phi,\psi)- g_{ij}\partial_{i}\partial^{i}F'(R,\phi,\psi))\nonumber\\
&+& (\frac{1}{2}  g_{ij}(G_{IJKL}G^{IJKL}) - R_{ij}(G_{IJKL}G^{IJKL})' + \partial_{i}\partial_{j}(G_{IJKL}G^{IJKL})'- g_{ij}\partial_{i}\partial^{i}(G_{IJKL}G^{IJKL})')\nonumber\\
&+& \varepsilon^{I'J'K'}C_{I'J'K'}(\frac{1}{2}  g_{ij}(G_{IJKL}G^{IJKL}) - R_{ij}(G_{IJKL}G^{IJKL})' + \partial_{i}\partial_{j}(G_{IJKL}G^{IJKL})'- g_{ij}\partial_{i}\partial^{i}(G_{IJKL}G^{IJKL})')\nonumber\\
&+& (G_{IJKL}G^{IJKL})(\frac{1}{2}  g_{ij}(\varepsilon^{I'J'K'}C_{I'J'K'}) - R_{ij}(\varepsilon^{I'J'K'}C_{I'J'K'})' + \partial_{i}\partial_{j}(\varepsilon^{I'J'K'}C_{I'J'K'})'- g_{ij}\partial_{i}\partial^{i}(\varepsilon^{I'J'K'}C_{I'J'K'})')\nonumber\\
&+& g_{ij}[\Big(\Sigma_{n=1}^{6}\Sigma_{m=0}^{n-6} \varepsilon^{J_{1}J_{2}..J_{2n-1}J_{2n}}F_{J_{1}J_{2}}..F_{J_{2n-1}J_{2n}}\varepsilon^{I_{1}I_{2}..I_{2m-1}I_{2m}}\partial^{I_{1}}\phi\partial^{I_{2}}\phi..\partial^{I_{2m-1}}\phi\partial^{I_{2m}}\phi \Big)\nonumber\\
&-&i^{m}\Big(\Sigma_{n=1}^{6}\Sigma_{m=0}^{6-n} \varepsilon^{J_{1}J_{2}..J_{2n-1}J_{2n}}F_{J_{1}J_{2}}..F_{J_{2n-1}J_{2n}}\varepsilon^{I_{1}I_{2}..I_{2m-1}I_{2m}}\bar{\psi} \gamma^{I_{1}}\partial^{I_{2}}\psi..\bar{\psi} \gamma^{2I_{m}-1}\partial^{I_{2m}}\psi \Big)\nonumber\\
&-&i^{m}\Big(\Sigma_{n=1}^{6}\Sigma_{m=0}^{6-n}\Sigma_{k=0}^{6-n-m} \varepsilon^{J_{1}J_{2}..J_{2n-1}J_{2n}}F_{J_{1}J_{2}}..F_{J_{2n-1}J_{2n}}\varepsilon^{I_{1}I_{2}..I_{2m-1}I_{2m}}\bar{\psi} \gamma^{I_{1}}\partial^{I_{2}}\psi..\bar{\psi} \gamma^{2I_{m}-1}\partial^{I_{2m}}\psi  \nonumber\\
&\times& \varepsilon^{I_{1}I_{2}..I_{2k-1}I_{2k}}\partial^{I_{1}}\phi\partial^{I_{2}}\phi..\partial^{I_{2k-1}}\phi\partial^{I_{2k}}\phi+...\Big)]=0 ,\label{s20}
\end{eqnarray}
where $'$ denotes the derivative with respect to the curvature $R$ and

\begin{eqnarray}
&&\Big(\frac{\partial}{\partial x_{I}}[\frac{\partial}{\partial (\partial_{I}\phi)}]-\frac{\partial}{\partial \phi} \Big) \nonumber\\
&\times& [\Big(\Sigma_{n=1}^{6}\Sigma_{m=0}^{n-6} \varepsilon^{J_{1}J_{2}..J_{2n-1}J_{2n}}F_{J_{1}J_{2}}..F_{J_{2n-1}J_{2n}}\varepsilon^{I_{1}I_{2}..I_{2m-1}I_{2m}}\partial^{I_{1}}\phi\partial^{I_{2}}\phi..\partial^{I_{2m-1}}\phi\partial^{I_{2m}}\phi \Big)\nonumber\\
&-&i^{m}\Big(\Sigma_{n=1}^{6}\Sigma_{m=0}^{6-n} \varepsilon^{J_{1}J_{2}..J_{2n-1}J_{2n}}F_{J_{1}J_{2}}..F_{J_{2n-1}J_{2n}}\varepsilon^{I_{1}I_{2}..I_{2m-1}I_{2m}}\bar{\psi} \gamma^{I_{1}}\partial^{I_{2}}\psi..\bar{\psi} \gamma^{2I_{m}-1}\partial^{I_{2m}}\psi \Big)\nonumber\\
&-&i^{m}\Big(\Sigma_{n=1}^{6}\Sigma_{m=0}^{6-n}\Sigma_{k=0}^{6-n-m} \varepsilon^{J_{1}J_{2}..J_{2n-1}J_{2n}}F_{J_{1}J_{2}}..F_{J_{2n-1}J_{2n}}\varepsilon^{I_{1}I_{2}..I_{2m-1}I_{2m}}\bar{\psi} \gamma^{I_{1}}\partial^{I_{2}}\psi..\bar{\psi} \gamma^{2I_{m}-1}\partial^{I_{2m}}\psi  \nonumber\\
&\times& \varepsilon^{I_{1}I_{2}..I_{2k-1}I_{2k}}\partial^{I_{1}}\phi\partial^{I_{2}}\phi..\partial^{I_{2k-1}}\phi\partial^{I_{2k}}\phi+...\Big)]\nonumber\\&&\Big(\frac{\partial}{\partial x_{I}}[\frac{\partial}{\partial (\partial_{I}\phi)}]-\frac{\partial}{\partial \phi} \Big) \nonumber\\
&\times& [G_{IJKL}G^{IJKL}+\varepsilon^{I'J'K'}C_{I'J'K'}G_{IJKL}G^{IJKL}+F(R,\phi,\psi)]=0 ,\label{s21}
\end{eqnarray}

\begin{eqnarray}
&&\Big(\frac{\partial}{\partial x_{I}}[\frac{\partial}{\partial (\partial_{I}\psi)}]-\frac{\partial}{\partial \psi} \Big)  \nonumber\\
&\times& [\Big(\Sigma_{n=1}^{6}\Sigma_{m=0}^{n-6} \varepsilon^{J_{1}J_{2}..J_{2n-1}J_{2n}}F_{J_{1}J_{2}}..F_{J_{2n-1}J_{2n}}\varepsilon^{I_{1}I_{2}..I_{2m-1}I_{2m}}\partial^{I_{1}}\phi\partial^{I_{2}}\phi..\partial^{I_{2m-1}}\phi\partial^{I_{2m}}\phi \Big)\nonumber\\
&-&i^{m}\Big(\Sigma_{n=1}^{6}\Sigma_{m=0}^{6-n} \varepsilon^{J_{1}J_{2}..J_{2n-1}J_{2n}}F_{J_{1}J_{2}}..F_{J_{2n-1}J_{2n}}\varepsilon^{I_{1}I_{2}..I_{2m-1}I_{2m}}\bar{\psi} \gamma^{I_{1}}\partial^{I_{2}}\psi..\bar{\psi} \gamma^{2I_{m}-1}\partial^{I_{2m}}\psi \Big)\nonumber\\
&-&i^{m}\Big(\Sigma_{n=1}^{6}\Sigma_{m=0}^{6-n}\Sigma_{k=0}^{6-n-m} \varepsilon^{J_{1}J_{2}..J_{2n-1}J_{2n}}F_{J_{1}J_{2}}..F_{J_{2n-1}J_{2n}}\varepsilon^{I_{1}I_{2}..I_{2m-1}I_{2m}}\bar{\psi} \gamma^{I_{1}}\partial^{I_{2}}\psi..\bar{\psi} \gamma^{2I_{m}-1}\partial^{I_{2m}}\psi  \nonumber\\
&\times& \varepsilon^{I_{1}I_{2}..I_{2k-1}I_{2k}}\partial^{I_{1}}\phi\partial^{I_{2}}\phi..\partial^{I_{2k-1}}\phi\partial^{I_{2k}}\phi+...\Big)]\nonumber\\&&\Big(\frac{\partial}{\partial x_{I}}[\frac{\partial}{\partial (\partial_{I}\psi)}]-\frac{\partial}{\partial \psi} \Big) \nonumber\\
&\times& [G_{IJKL}G^{IJKL}+\varepsilon^{I'J'K'}C_{I'J'K'}G_{IJKL}G^{IJKL}+F(R,\phi,\psi)]=0 ,\label{s22}
\end{eqnarray}

\begin{eqnarray}
&&\Big(\frac{\partial}{\partial x_{I}}[\frac{\partial}{\partial (\partial_{I}A_{J})}]-\frac{\partial}{\partial A_{I}}  \Big) \nonumber\\
&\times& [\Big(\Sigma_{n=1}^{6}\Sigma_{m=0}^{n-6} \varepsilon^{J_{1}J_{2}..J_{2n-1}J_{2n}}F_{J_{1}J_{2}}..F_{J_{2n-1}J_{2n}}\varepsilon^{I_{1}I_{2}..I_{2m-1}I_{2m}}\partial^{I_{1}}\phi\partial^{I_{2}}\phi..\partial^{I_{2m-1}}\phi\partial^{I_{2m}}\phi \Big)\nonumber\\
&-&i^{m}\Big(\Sigma_{n=1}^{6}\Sigma_{m=0}^{6-n} \varepsilon^{J_{1}J_{2}..J_{2n-1}J_{2n}}F_{J_{1}J_{2}}..F_{J_{2n-1}J_{2n}}\varepsilon^{I_{1}I_{2}..I_{2m-1}I_{2m}}\bar{\psi} \gamma^{I_{1}}\partial^{I_{2}}\psi..\bar{\psi} \gamma^{2I_{m}-1}\partial^{I_{2m}}\psi \Big)\nonumber\\
&-&i^{m}\Big(\Sigma_{n=1}^{6}\Sigma_{m=0}^{6-n}\Sigma_{k=0}^{6-n-m} \varepsilon^{J_{1}J_{2}..J_{2n-1}J_{2n}}F_{J_{1}J_{2}}..F_{J_{2n-1}J_{2n}}\varepsilon^{I_{1}I_{2}..I_{2m-1}I_{2m}}\bar{\psi} \gamma^{I_{1}}\partial^{I_{2}}\psi..\bar{\psi} \gamma^{2I_{m}-1}\partial^{I_{2m}}\psi  \nonumber\\
&\times& \varepsilon^{I_{1}I_{2}..I_{2k-1}I_{2k}}\partial^{I_{1}}\phi\partial^{I_{2}}\phi..\partial^{I_{2k-1}}\phi\partial^{I_{2k}}\phi+...\Big)]\nonumber\\&&\Big(\frac{\partial}{\partial x_{I}}[\frac{\partial}{\partial (\partial_{I}A_{J})}]-\frac{\partial}{\partial A_{I}}  \Big) \nonumber\\
&\times& [G_{IJKL}G^{IJKL}+\varepsilon^{I'J'K'}C_{I'J'K'}G_{IJKL}G^{IJKL}+F(R,\phi,\psi)]=0 ,\label{s23}
\end{eqnarray}

For a 4-dimensional universe, we have $R=12 H^{2} + 6 \dot{H}$, where $H=\frac{\dot{a}}{a}$ is the Hubble parameter and $a$ is the scale factor of the universe. Assuming that all fields depend only on time,  and  using the relation between scalars, gauge fields, curvatures and strings in equation (\ref{s2}), we can solve equations (\ref{s20}, \ref{s21}, \ref{s22} and \ref{s23}) simultanously and obtain:

 \begin{eqnarray}, 
a(t) &\approx& e^{-\int dt Z(t)},\nonumber\\ Z(t) &\approx&  \Sigma_{n=1}^{6}\beta_{n}t_{s}^{2n}(1-\frac{t}{t_{s}})^{2n}\Big(1+\Sigma_{m=0}^{6-n}\alpha_{m}ln^{2m}(t_{s}(1-\frac{t}{t_{s}}))[1+\Sigma_{k=0}^{12-2n-2m}\lambda_{k}e^{2\gamma_{k}kt_{s}(1-\frac{t}{t_{s}})}]\Big)\nonumber\\ \nonumber\\ \nonumber\\ \phi(t) &\approx& (\psi)^{\frac{1}{2}} \approx \Sigma_{n=1}^{6}\beta_{n}t_{s}^{2n}(1-\frac{t}{t_{s}})^{2n}\Sigma_{m=0}^{6-n}\alpha_{m}ln^{-2m}(\frac{t}{t_{s}})\times Z^{\frac{1}{4}}(t) \nonumber\\ \nonumber\\ \nonumber\\ A^{I}(t) &\approx&  \varepsilon^{I}\Sigma_{m=1}^{6}\alpha_{m}(t_{s}(1-\frac{t}{t_{s}})^{2m}\Sigma_{k=0}^{6-m}\lambda'_{k}e^{-2\gamma'_{k}k(\frac{t}{t_{s}})}\times ln (Z(t))\label{s24}
\end{eqnarray}
where $t_{s}$ is time of collidision between the two 11-dimensional manifolds. This time is much larger than the present age of the universe ($t_{s}\lll 13.5 Gyr$). The above result shows the coincidence with the birth of the universe ($t=0$) on the 11-dimensional manifold.  The scale factor grows with time, while scalars, fermions and gauge fields decrease,  and tend to zero at ($t=t_{s}$). This is because with the passage of time, the two 11-dimensional manifolds come close to each other and all fields which live on the 3-dimensional manifold between them dissolve into the  11-dimensional manifolds. By the disappearance of these fields, gravitons  and $F(R)$-gravity emerge that lead to the expansion and inflation of the  universe (See Figure 7).

\begin{figure*}[thbp]
	\begin{center}
		\begin{tabular}{rl}
			\includegraphics[width=8cm]{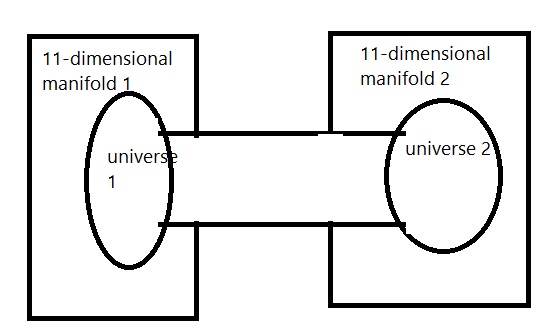}
		\end{tabular}
	\end{center}
	\caption{  Emergence of 4-dimensional universe on 11-dimensional manifold in 11+3-dimensional space-time. }
\end{figure*}

We can test our model by calculating the magnitude of the slow-roll parameters and the tensor-to-scalar ratio ($R_{tensor-scalar}$)
defined in \cite{da} and compare with previous predictions:

\begin{eqnarray}
&& H=\frac{\dot{a}}{a}= -\dot{Z} \Rightarrow \nonumber \\  && \varepsilon=-\frac{\dot{H}}{H^{2}}=\frac{\ddot{Z}}{(\dot{Z})^{2}}\nonumber \\ && \nonumber \\ &&\nonumber \\ && \eta=-\frac{\ddot{H}}{2H\dot{H}}=\frac{\dddot{Z}}{2\dot{Z}\ddot{Z}} \nonumber \\ && \nonumber \\ &&\nonumber \\ && R_{tensor-scalar}= 16\varepsilon=16\frac{\ddot{Z}}{(\dot{Z})^{2}}
\label{kj1}
\end{eqnarray}

Obviously, during the inflationary epoch, the age of the universe (t) is very smaller with respect to the time of collision between manifolds ($t_{s}$) and, as a result, ($(t_{s}-t)^{-n}\ll 1$) where $n$ is an integer.  Using equation (\ref{kj1}) and expanding the functions ($e^{O}$ and $ln$) by applying the Taylor series, we can derive  the following results:

\begin{eqnarray}
&& 0\ll t\ll t_{s}\Rightarrow   \eta \approx \Sigma \beta_{n} (t_{s}-t)^{-2n}(ln^{-2n}(t_{s}(1-\frac{t}{t_{s}}))+ O(\text{smaller terms}))\ll 1 \nonumber\\&& \varepsilon\approx \Sigma \alpha_{m}(t_{s}-t)^{-2m}(e^{-2\gamma_{m}mt_{s}(1-\frac{t}{t_{s}})}+ O(\text{smaller terms}))\ll 1 \nonumber\\&& \Rightarrow  R_{tensor-scalar}\ll 1\label{kj2}
\end{eqnarray}

This equation indicates that   the values of the slow-roll parameters and the tensor-to-scalar ratio are very much smaller than the ones which are  consistent with the predictions of the experiments in ref \cite{pl}. Thus, this type of $F(R,\phi,\psi)$-gravity which is produced as due to the interaction of two 11-dimensional manifolds, creates the correct values for the experimental  parameters and can describe cosmological events.

\section{Summary and conclusion }\label{o4}
Motivated by the successes of modified theories of gravity such as $F(R)$ gravity, in this paper, we have investigated how we could make $F(R)$ gravity anomaly-free. To achieve this, we considered the origin of the action of $F(R)$-gravity on an (11+3)-dimensional space-time. However, we found that this was not anomaly-free. So, we proposed a theory which lives on an (11+3)-dimensional manifold with two 11-dimensional manifolds and one
3-dimensional manifold. We showed that our universe is a part of one  11-dimensional manifold which is connected
with the other 11-dimensional manifold by an extra 3-dimensional manifold. The two 11-dimensional manifolds interact
with each other via exchanging fields which move along the 3-dimensional manifold. These fields are the main cause
for the appearance of F(R)-gravity in the four dimensional spacetime. This is because with the passage of time, the two 11-dimensional manifolds come close to each other and all ﬁelds which live on the 3-dimensional manifold between them dissolve into the 11-dimensional manifolds. As these fields disappear, gravitons and F(R)-gravity emerge that lead to the expansion and inﬂation of the universe.  Hence, the interaction between the two 11-dimensional manifolds has a direct efect on the evolution of the universe during the inflationary era. We have  calculated the scale factor of the inflationary universe, and examined our model against WMAP and Planck experiments. We have obtained the slow-roll parameters and shown that the tensor-scalar ratio is much  smaller than  unity ($R_{tensor-scalar}\lll 1$), which is consistent with experimental observations.

\section*{Acknowledgments}
\noindent Aroonkumar Beesham acknowledges that this work is based on the research supported wholly / in part by the National Research Foundation of South Africa (Grant Numbers: 118511). The work of Kazuharu Bamba has been supported in part by the JSPS KAKENHI Grant Number JP 25800136 and Competitive Research Funds for Fukushima University Faculty (19RI017).


\end{document}